# Transmitter IQ Skew Calibration using Direct Detection

**Wing Chau Ng, Xuefeng Tang, Zhuhong Zhang**
*Huawei Technologies Canada, 303 Terry Fox Dr., Ottawa, Canada*
*wing.chau.ng@huawei.com*

**Abstract:** We propose a transmitter skew calibration based on direct detection of coherent signals with estimation errors of ±0.2ps, providing a reliable, accurate and low-cost scheme to calibrate skew for coherent transceivers. © 2020 The Author(s)

## 1. Introduction

High symbol rates and high-order modulation formats are important to realize spectrally efficient dense wavelength division multiplexing transmission in optical networks. For high Baud transmissions (400Gb/s or beyond), a sub-picosecond timing misalignment (skew) between the in-phase (I) and quadrature (Q) tributaries of the IQ modulator at transmitter (Tx) may degrade significantly system performance [1]. Receiver (Rx) digital signal processing (DSP) can be designed to compensate Tx IQ skew to some extent [2], but the implementation requires high complexity and power consumption. Also, channel impairments such as fast polarization rotation, polarization mode dispersion and laser phase noise degrade its compensation performance. Therefore, skew calibration is indispensable at the Tx side [1]. Skew can be calibrated via coherent detection (cohD) with DSP [3]. To avoid additional IQ imbalance and device impairments introduced by integrated coherent receiver (ICR) and analog-to-digital converters (ADCs) that complicate Tx characterization and introduce extra cost, direct detection (DD) is preferred.

Several proposed low-bandwidth DD approaches require IQ phase off from quadrature [4, 5] and may require additional electronics to generate RF (radio frequency) tones [5]. Changing phase bias affects I and Q bias points, resulting in device responses different from the operating condition. It also takes time to stabilize at new bias points with manual operations or special algorithms, causing inconvenience in calibration. Some high-bandwidth approaches require the detection of image spectral power [6, 7], but their accuracy is degraded by the IQ phase error. A genetic algorithm is proposed [8], but still with estimation errors of ±0.5ps. Instead of using these techniques inferring skew, the most straightforward and reliable way is to present the coherent signal constellation detected by a single photodetector (PD), and estimate imbalance parameters via standard algorithms. Such a DD constellation-based method tells what exactly the signal experiences.

The recent advancement in DD enables the recovery of coherent signals via direct detection, with the aid of an additional frequency tone to beat with the signal at a PD) [9]. Contrary to transmission scenario, the carrier-to-signal ratio (CSPR) and signal-signal beat interference (SSBI) are not critical for Tx calibration or monitoring purposes thanks to the absence of transmission impairments. Therefore, SSBI cancellation is not required, simplifying calibration algorithms. By taking these advantages, in this paper, we demonstrate a DD constellation-based Tx IQ skew calibration using a 34GBaud 16QAM signal. Our experimental result yields estimation errors of +/-0.2ps, suggesting that our technique is reliable, accurate and cost-saving to measure pure Tx IQ skew.

## 3. Principle and experiment

Direct detection does not recover coherent quadrature amplitude modulation (QAM) signals because phase information is lost during the squaring operation upon intensity detection. One can reconstruct a complex field from the photocurrent given that the real and the imaginary part of the signals are correlated, i.e., beat the QAM signal with an additional laser source with a frequency shift larger than the symbol rate. In our experiment, as shown in Fig. 1, there are two feedback loops in red dashed boxes. The first is the automatic bias control, available in contemporary transmitter products. The second feedback loop is proposed in this work for skew estimation during the factory calibration stage: A 34GBaud 16QAM signal was pre-emphasized by the measured S21 response, without SSBI cancellation. Its real and imaginary parts generated by digital-to-analog converters (DACs) were amplified by the RF drivers. The output optical signal (-16.5dBm) from an IQ modulator under test and another laser source (-3dBm), roughly 20GHz away from the signal carrier, were detected together by a 37GHz bandwidth PD. The 40mV peak-to-peak PD electrical output was sampled at 160Gsa/s by a digital sampling oscilloscope (DSO).

For offline processing, the digitized PD output was first resampled to 8 samples per symbol which contains a DC term, a laser-signal beat term (the real part of the mixed signal) and a SSBI term. Since there is no transmission impairment (e.g. optical amplifier noise), the recovered coherent signal quality is good enough to calibrate Tx without SSBI cancelation. After DC removal, a traditional algorithm was used for field reconstruction: the corresponding imaginary of the mixed signal was obtained via frequency-domain Hilbert transform and then combined with the real part, resulting in a bandpass 16QAM signal with a frequency shift of around 20GHz. The frequency offset was estimated by the well-known maximization of the $4^{th}$ power of signal spectrum, and was then compensated to generate a baseband 16QAM signal. An anti-aliasing filter suppressed the out-of-band spectral residuals (SSBI and tones). As a comparison, Kramers–Kronig (KK) receiver algorithm [9] was also used to replace the traditional algorithm for field reconstruction. After data synchronization, carrier recovery and Wiener filter were applied to recover the original 16QAM signal. The IQ skew can be estimated via filter tap values. To verify our DD calibration, the modulator output was sent to a cohD setup for skew characterization, using the same DSP algorithms without field reconstruction.

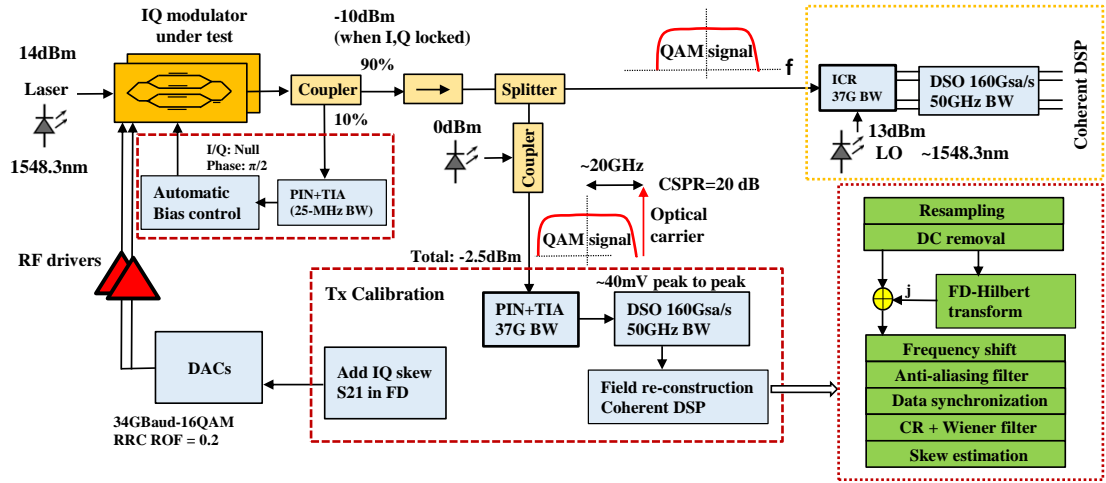

Fig. 1. Experimental testbed for direct-detection based factory calibration for Tx IQ modulators.

## 3. Experimental results

To evaluate the performance of the proposed skew estimation, IQ skews from -8ps to 9ps were intentionally introduced as a linear phase in frequency domain using DAC. The optical output from the modulator was partially fed into a PD to perform a DD-based Tx calibration, shown in the lower dashed red box in Fig.1, as well as into the standard cohD setup shown in the right dotted orange box for verification. Since both techniques recover the signal constellation, the signal's error vector magnitudes (EVM) were calculated using the known transmitted data.

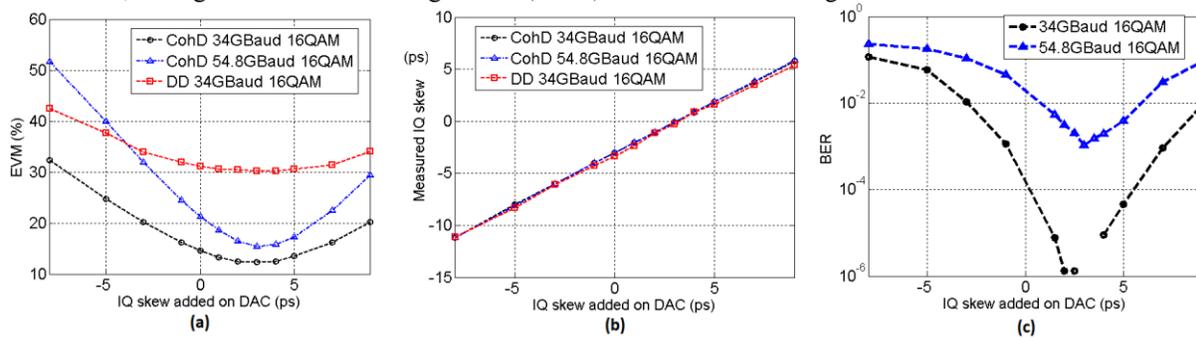

Fig. 2. Comparison of DD-based and coherent-detection (CohD)-based skew measurement. (a) EVM versus IQ skew added on Tx DAC. (b) Measured IQ skew versus IQ skew added on Tx DAC. (c) BER of 34GBaud and 54.8GBaud 16QAM measured by our Rx DSP versus IQ skew added on Tx DAC.

Fig. 2a shows the EVM values versus the skew added on the transmitted signal, shown that our DD method (squared, red dashed) has the same minimum EVM location as the conventional coherent detection method (circle, black dashed). The experiment was repeated using a 54.8GBaud 16-QAM (triangle, blue dash-dotted). The result verified with the previous two cases, with a large slope change because of higher sensitivity to transmitter skew at higher Baud. Fig. 2b shows the estimated skews of the both DD and cohD methods. We can see that the DD results aligned with the CohD results. Note that the CohD method may suffer from uncertainties due to the imbalance and device responses of ICR and ADCs, as well as the absence of laser frequency offset and SOP rotation [3]. The results using KK algorithm are almost the same those using the traditional algorithm, and therefore are not shown in Fig. 2b for clarity. The curves are linear with unity slope and a y-intercept at -3.0ps, suggesting that the intrinsic skew of the IQ modulator under test should be at -3.0 ps.

To verify the DD-based measurement, Fig. 2c shows the bit error rate (BER) measured by coherent Rx DSP (without Tx skew compensation) versus IQ skew added on both 34Gbaud and 54.8GBaud 16QAM signals. The BER has a minimum at 3.0ps, i.e., one has to add 3.0ps on Tx signal to compensate the intrinsic skew of -3.0ps, which aligned with the above estimate. Fig. 3a shows the histograms of estimated skew using the two field reconstruction algorithms based on 300 experimental results. The traditional algorithm (green) and the KK receiver algorithm (dark) yield sample means of -2.968ps and -2.975ps, respectively, with sample standard deviations of 0.0708ps and 0.0695ps respectively. Based on the sampling distributions in Fig. 3a, a single-shot estimate (without adding skew) from both algorithms yield values between -3.2ps and -2.8ps, i.e., the estimation errors are ±0.2ps. Fig. 3b shows the simulated OSNR penalty versus Tx skew for a skew-sensitive 800Gb/s system. The OSNR penalty increases more than 0.5 dB when the Tx skew is larger than 0.45ps. It is thus important to calibrate skew accurately on the Tx side. The proposed calibration introduces OSNR penalty less than 0.1 dB in a 800Gb/s system.

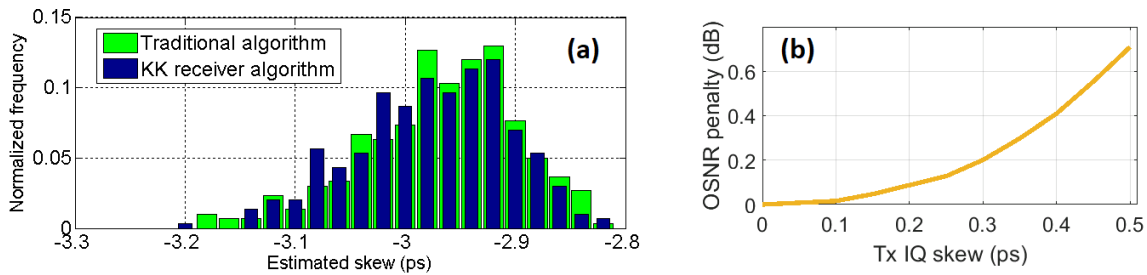

Fig. 3 (a) Histograms (with 300 samples) of estimated skew using two field reconstruction algorithms. Green: traditional algorithm, Dark blue: KK receiver algorithm. Sample size is 300. (b) OSNR penalty versus Tx IQ skew of a 800Gb/s, 138GBaud DP-16QAM back-to-back system (ROF = 0.2, BER = 1e-2).

## 3. Conclusion

We proposed a transmitter IQ skew calibration by recovering coherent QAM signals using a wideband photodetector and a high speed ADC, with estimation errors of ±0.2ps, introducing an OSNR penalty less than 0.1 dB in a 800Gb/s system. Compared to coherent detection, our proposed DD-based Tx calibration does not require the conditions such as sufficient frequency offset between Tx and Rx, and polarization alignment [4]. Contrary to the previously proposed techniques [4, 5], our scheme does not require IQ phase bias detuning, and additional RF electronics is not required. Contrary to transmission scenario, DD-based calibration DSP does not require SSBI cancellation nor carefully choose CSPR, simplifying algorithms for constellation recovery. Other Tx parameters such as quadrature phase error can also be estimated. As ICR is not required and the constellation is recovered, this scheme is considered as a reliable, low-cost and accurate Tx skew calibration approach and could serve as a hardware foundation to prototype transmitter self-calibration, in-line digital pre-distortion and transmitter monitoring.